\documentclass[prl,aps,floats,twocolumn]{revtex4}

\usepackage[dvips]{graphicx}
\usepackage{amssymb}
\usepackage{amsmath}

\usepackage{color}

\begin{document}

\title{The general theory of porcupines, perfect and imperfect}

\author{Latham Boyle}

\affiliation{Perimeter Institute for Theoretical Physics, Waterloo,
  Ontario, Canada}

\date{August 2010}
                            
\begin{abstract}
  Porcupines are networks of gravitational wave detectors in which the
  detectors and the distances between them are short relative to the
  gravitational wavelengths of interest.  Perfect porcupines are
  special configurations whose sensitivity to a gravitational plane
  wave is independent of the propagation direction or polarization of
  the wave.  I develop the theory of porcupines, including the optimal
  estimator $\hat{h}^{ij}$ for the gravitational wave field; useful
  formulae for the spin-averaged and rotationally-averaged SNR$^{2}$;
  and a simple derivation of the properties of perfect porcupines.  I
  apply these results to the interesting class of ``simple''
  porcupines, and mention some open problems.
\end{abstract}
\maketitle 

%\section{Introduction}

This decade, we hope and expect that gravitational waves will be
directly detected for the first time; this will mark the beginning of
what promises to be a long and fruitful era of gravitational wave
astronomy.  One can only guess what wonders will be revealed when this
new window onto the universe is flung fully open.  As attention shifts
from gravitational wave {\it detection} to gravitational wave {\it
  astronomy}, so will it shift from individual gravitational wave {\it
  detectors} to networks of multiple detectors that that function
together as gravitational wave {\it telescopes}.  (For previous work
on gravitational wave networks, see \cite{GurselTinto,
  JohnsonMerkowitz, CerdonioEtAl, WenChen} and references therein.)
There is an important regime in which the individual detectors and the
distances between them are short relative to the gravitational
wavelengths of interest, so that the arms of the various detectors in
the network may be thought of as emanating from nearly the same point
in space, like the fanned quills of a frightened porcupine.  In a
previous paper \cite{porcupine}, such ``porcupines'' were motivated by
recent ideas and developments in gravitational wave detection
\cite{LIGO, DimopoulosEtAl, HarmsEtAl, YuTinto, HohenseeEtAl}; but it
is worth reframing the motivation in more general terms: one may
suspect that porcupines will play an important role in gravitational
wave astronomy for the simple reason that astrophysical gravitational
wavelengths are typically very {\it long}.  One will often be
interested in gravitational wavelengths that are longer than the
practical size of one's network ({\it e.g.}\ the diameter of the
Earth, or the Solar System).

Ref.~\cite{porcupine} focused on ``perfect porcupines'': special
configurations with the defining property that their sensitivity to an
incident gravitational plane wave is independent of the propagation
direction or polarization of the wave.  (A number of other nice
properties follow automatically once this condition is satisfied.)
Here I develop a more natural and general formalism for handling
porcupines (perfect or imperfect).  This formalism yields, for an
arbitrary imperfect porcupine: (i) the minimum-variance unbiased
estimator $\hat{h}^{ij}$ for the gravitational wave field; and (ii)
useful expressions for the spin-averaged and rotationally-averaged
SNR$^{2}$.  It also yields a simple derivation of the properties of
perfect porcupines.  I apply these results to the interesting class of
``simple'' porcupines (defined below) and discuss open problems.

\section{General Formalism}
\label{formalism}

To fix notation, my fourier conventions are
\begin{equation}
  g(t)\!=\!\!\int_{-\infty}^{\infty}\!\!\!df\,\tilde{g}(f){\rm e}^{-2\pi ift},
  \quad
  \tilde{g}(f)\!=\!\!\int_{-\infty}^{\infty}\!\!\!dt\,g(t){\rm e}^{+2\pi ift}.
\end{equation}
The lower case latin indices $\{i,j,k,l,m,n\}$ label the 3 spatial
directions: $i,j,k,l,m,n=1,2,3$.  The upper case latin indices
$\{A,B\}$ label the 2 gravitational wave polarizations: $A,B=1,2$.
The lower case greek indices $\{\alpha,\beta\}$ label the $N$
detectors in the network: $\alpha,\beta=1,\ldots,N$.  Indices in
square braces $[\ldots]$ are symmetrized.  I use the Einstein
summation convention: repeated indices (one upper, one lower) are
summed.  I use hats in two different ways: (i) to denote variables
with a random noise contribution (such as estimators); and (ii) to
denote unit 3-vectors.  I have tried to make the context clear enough
to avoid confusion of these two meanings.

A gravitational wave on Minkowski space is described in
transverse-traceless gauge \cite{MTW} by the line element
\begin{equation}
  \label{metric}
  ds^{2}=-dt^{2}+[\delta_{ij}+2h_{ij}]dx^{i}dx^{j}.
\end{equation}
In this metric, worldlines with $\vec{x}={\rm constant}$ are
geodesics; along such worldlines, the proper time $\tau$ coincides
with the coordinate time $t$.  When gravitational waves reach us from
a distant astronomical source, they appear to us as plane waves
travelling in the $\hat{n}$ direction
\begin{equation}
  h^{ij}(\lambda)=P^{ij}_{A}(\hat{n})h^{A}(\lambda)
\end{equation}
where the polarization waveforms $h_{}^{A}(\lambda)$ are arbitrary
functions of $\lambda\equiv t-\hat{n}\cdot\hat{x}$, and the
polarization tensors $P^{ij}_{A}(\hat{n})$ form an orthonormal basis
on the 2-dimensional space of symmetric, transverse, traceless
$3\times3$ matrices:
\begin{subequations}
  \begin{eqnarray}
    P^{ij}_{A}(\hat{n})-P^{ji}_{A}(\hat{n})&=&0, \\
    \hat{n}_{i}P^{ij}_{A}(\hat{n})&=&0, \\
    \delta_{ij}P^{ij}_{A}(\hat{n})&=&0, \\
    P^{B}_{ij}(\hat{n})^{\ast}P^{ij}_{A}(\hat{n})&=&\delta_{\;\;\,A}^{B}\,.
  \end{eqnarray}
\end{subequations}
Let us imagine that all of the detectors are situated at $\vec{x}=0$.
The estimator $\hat{h}^{\alpha}$ represents the measured output of
detector $\alpha$, a sum of genuine gravitational wave signal
$h^{\alpha}(t)$ and noise $\hat{n}^{\alpha}(t)$:
\begin{equation}
  \label{s_alpha}
  \hat{h}^{\alpha}(t)=h^{\alpha}(t)+\hat{n}^{\alpha}(t).
\end{equation}
If we assume the network's response is linear and time-translation
invariant, we can write
\begin{equation}
  h^{\alpha}(t)=\int_{-\infty}^{+\infty}dT\,W^{\alpha}_{ij}(T)h^{ij}(t-T).
\end{equation}
Since $h^{\alpha}(t)$ is real and $h^{ij}(t)$ is real and symmetric,
$W^{\alpha}_{ij}(t)$ is also real and symmetric.  We model the noise
$\hat{n}^{\alpha}(t)$ as stationary and gaussian, with zero mean, so
it is characterized by its correlation function
$C^{\alpha}_{\;\;\beta}(T)$ or, equivalently, its spectral density
$S^{\alpha}_{\;\;\beta}(f)= \tilde{C}^{\alpha}_{\;\;\beta}(f)$:
\begin{subequations}
  \begin{eqnarray}
    C^{\alpha}_{\;\;\beta}(T)&=&
    \langle\hat{n}^{\alpha}(t+T)\hat{n}_{\beta}(t)\rangle, \\
    \delta(f-f')S^{\alpha}_{\;\;\beta}(f)&=&
    \langle\hat{n}^{\alpha}(f)\hat{n}_{\beta}^{\ast}(f')\rangle.
 \end{eqnarray}
\end{subequations}
$S^{\alpha}_{\;\;\beta}(f)$ induces a natural inner product on the
space of signals (or noise) in the network:
\begin{equation}
  \big(p\big|q\big)=\int_{-\infty}^{\infty}df\,
  \tilde{p}_{\alpha}^{\,\ast}(f)[S^{-1}(f)]^{\alpha}_{\;\;\beta}
  \,\tilde{q}_{}^{\,\beta}(f).
\end{equation}
Using matched filtering, a given gravitational wave signal $h^{ij}(t)$
can be detected by such a network with expected signal-to-noise 
(SNR) given by
\begin{subequations}
  \label{SNR}
  \begin{eqnarray}
    {\rm SNR}^{2}\!=\!(h|h)\!\!&\!=\!&\!\!\!
    \int_{-\infty}^{\infty}\!\!df\tilde{h}_{\alpha}^{\ast}(f)
    [S^{-1}(f)]^{\alpha}_{\;\;\beta}\,\tilde{h}_{}^{\beta}(f) \\
    \!\!&\!=\!&\!\!\!\int_{-\infty}^{\infty}\!\!df 
    \tilde{h}_{ij}^{\ast}(f)K^{ij}_{\;\;\;kl}(f)\tilde{h}_{}^{kl}(f)\quad
 \end{eqnarray}
\end{subequations}
where we have introduced a kernel
\begin{equation}
  \label{def_K}
  K^{ij}_{\;\;\;kl}(f)\equiv \tilde{W}_{\alpha}^{ij}(f)^{\ast}
  [S^{-1}(f)]^{\alpha}_{\;\;\beta}\,\tilde{W}^{\beta}_{kl}(f)
\end{equation}
with the following properties
\begin{equation}
  \label{K_constraints}
  K^{ij}_{\;\;\;kl}(f)\!=\!
  K^{[ij]}_{\;\;\;\;\;[kl]}(f)\!=\!
  K^{kl}_{\;\;\;ij}(f)^{\ast}\!=\!
  K^{ij}_{\;\;\;kl}(-f)^{\ast}.
\end{equation}
If a gravitational wave signal (which depends on various parameters
$\mu^{a}$) is detected, and the likelihood function may be
approximated as gaussian $\propto{\rm exp}[-(1/2)\mu^{a}
\Gamma_{ab}\mu^{b}]$ near its peak, then the expected inverse
covariance matrix is the Fisher information matrix, given by
\begin{equation}
  \Gamma_{ab}=\Big(\frac{\partial h}{\partial\mu^{a}}\Big|
  \frac{\partial h}{\partial\mu^{b}}\Big).
\end{equation}
Now consider the angular resolution of this network.  Let us define an
orthonormal triad consisting of $\hat{n}$ and two additional unit
vectors $\hat{m}_{\bar{\mu}}$ ($\bar{\mu}=1,2$).  Under a rotation
around the direction $\hat{m}_{\bar{\mu}}$ by an infinitessimal angle
$\theta^{\bar{\mu}}$, $h^{ij}(t)$ transforms as $h^{ij}(t)\to
R^{i}_{\;k}R^{j}_{\;\,l}h^{kl}(t)$ where
$R^{i}_{\;j}\approx\delta^{i}_{\;j}-\epsilon^{i}_{\;jk}
\hat{m}_{\bar{\mu}}^{k}\theta^{\bar{\mu}}$.  So the angular part of
$\Gamma_{ab}$ is given by
\begin{equation}
    \Gamma_{\bar{\mu}\bar{\nu}}\!=\!
    \Big(\frac{\partial h}{\partial\theta^{\bar{\mu}}}
    \Big|\frac{\partial h}{\partial\theta^{\bar{\nu}}}\Big)\!=\!\!
    \int_{-\infty}^{\infty}\!\!df \tilde{h}_{ij}^{\ast}(f)
    L^{ij}_{\bar{\mu}\bar{\nu}kl}(f)\tilde{h}_{}^{kl}(f)
\end{equation}
where we have defined
\begin{subequations}
  \begin{eqnarray}
   \hat{L}^{ij}_{\bar{\mu}\bar{\nu}i'\!j'\!}(f)&\equiv&
    4K^{ik}_{\;\;\;i'\!k'\!}(f)\epsilon^{j}_{\;kl}\epsilon_{j'}^{\;\;k_{}'l_{}'}
    \hat{m}_{\bar{\mu}}^{l}\hat{m}_{\bar{\nu}l'}^{} \\
    L^{ij}_{\bar{\mu}\bar{\nu}i'\!j'\!}(f)&\equiv&
    \hat{L}^{[ij]}_{\bar{\mu}\bar{\nu}[i'\!j'\!]}(f).
  \end{eqnarray}
\end{subequations}
For fixed $\bar{\mu}$ and $\bar{\nu}$,
$L^{ij}_{\bar{\mu}\bar{\nu}kl}(f)$ has properties exactly akin to
those in (\ref{K_constraints}).  The ``off-diagonal'' terms in the
Fisher matrix, with one index $\bar{\mu}$ corresponding to an angular
parameter $\theta^{\bar{\mu}}$, and the other index $\chi$
corresponding to any other (non-angular) parameter is
\begin{equation}
  \Gamma_{\chi\bar{\mu}}=2\!\!\int_{-\infty}^{\infty}\!\!\!\!\!df
  \frac{\partial\tilde{h}_{\!A}^{\ast}(f)}{\partial\chi}
  P^{A}_{ij}(\hat{n})^{\ast}K^{ij}_{\;\;\,kl}(f)
  \epsilon^{k}_{\;\,rs}\hat{m}_{\bar{\mu}}^{r}h^{ls}(f).
\end{equation}

\section{Optimal estimator for $h^{ij}$}

We want to find the minimum-variance unbiased estimator
$\hat{h}^{ij}(f)$ and its covariance.  Since
$\tilde{h}_{}^{\alpha}(f)$ is linearly related to $\tilde{h}^{ij}(f)$
[$\tilde{h}^{\alpha}(f)=\tilde{W}^{\alpha}_{ij}(f)\tilde{h}^{ij}(f)$],
$\hat{h}^{ij}(f)$ is linearly related to $\hat{h}^{\alpha}(f)$:
\begin{equation}
  \label{def_V}
  \hat{h}^{ij}(f)=\tilde{V}^{ij}_{\;\alpha}(f)\hat{h}^{\alpha}(f).
\end{equation}
It is stationary and gaussian, with covariance
\begin{equation}
  \label{def_Sijkl}
  \langle\delta\hat{h}^{ij}(f)
  \delta\hat{h}_{kl}^{\ast}(f')\rangle
  =S^{ij}_{\;\;\;kl}(f)\delta(f-f')
\end{equation}
where $\delta\hat{h}^{ij}\equiv\hat{h}^{ij}-\tilde{h}^{ij}$.  Our task
is to express $\tilde{V}^{ij}_{\;\alpha}$ and $S^{ij}_{\;\;\;kl}$ in
terms of the given tensors $\tilde{W}^{\alpha}_{ij}$ and
$S^{\alpha}_{\;\;\beta}$ that define the porcupine.

The estimator is unbiased if $\langle\hat{h}^{ij}\rangle
=\tilde{h}^{ij}$, and hence $\tilde{h}^{ij}=\tilde{V}^{ij}_{\;\alpha}
\tilde{W}^{\alpha}_{kl} \tilde{h}^{kl}$.  Since this must hold for
arbitrary $\tilde{h}^{ij}$, we learn that
\begin{equation}
  \label{VW_constraint}
  \tilde{V}^{ij}_{\;\alpha}\tilde{W}^{\alpha}_{kl}=I^{ij}_{\;\;\,kl}
\end{equation}
where
\begin{equation}
  \label{def_Iijkl}
  I^{ij}_{\;\;\,kl}\equiv\delta^{[i}_{\;\;[k}\delta^{j]}_{\;\;\,l]}
\end{equation}
is the identity operator on the 6-dimensional vector space of
$3\times3$ symmetric matrices.  The estimator $\hat{h}_{ij}$ only
exists if the porcupine tensor $\tilde{W}^{\alpha}_{ij}$ has a
left-inverse $\tilde{V}^{ij}_{\;\alpha}$ in the sense of
(\ref{VW_constraint}).  In a porcupine consisting of $N$ detectors: if
$N<6$, $\tilde{V}^{ij}_{\;\alpha}$ does not exist; if $N=6$ then, when
$\tilde{V}^{ij}_{\;\alpha}$ exists, it is specified uniquely by
(\ref{VW_constraint}); and if $N>6$ then, when
$\tilde{V}^{ij}_{\;\alpha}$ exists, it is {\it not} specified uniquely
by (\ref{VW_constraint}).  In this last case, the residual ambiguity
in $\tilde{V}^{ij}_{\;\alpha}$ may be fixed by using the method of
Lagrange multipliers to minimize the total variance
$I^{ij}_{\;\;\;kl}S^{kl}_{\;\;\;ij}$ of the estimator $\hat{h}_{ij}$,
subject to the constraint (\ref{VW_constraint}); for help, see {\it
  e.g.}\ Appendix D in \cite{BoyleKesden}.  In this way we find
\begin{equation}
  \label{V_minimum_variance}
  \tilde{V}^{ij}_{\;\alpha}=(D^{-1})^{ij}_{\;\;\;kl}
  \tilde{W}^{kl\ast}_{\,\beta}(S^{-1})^{\beta}_{\;\;\alpha}
\end{equation}
and
\begin{equation}
  \label{S_minimum_variance}
  S^{ij}_{\;\;\;kl}=(D^{-1})^{ij}_{\;\;\;kl}
\end{equation}
where we have defined the tensor $D$ and its inverse 
$D^{-1}$\begin{subequations}
  \label{def_D}
  \begin{eqnarray}
    \label{D}
    &D^{ij}_{\;\;\,kl}\equiv\tilde{W}^{ij\ast}_{\,\alpha}
    (S^{-1})^{\alpha}_{\;\;\beta}\tilde{W}^{\beta}_{kl},& \\
    \label{D_inverse}
    &D^{ij}_{\;\;\;mn}(D^{-1})^{mn}_{\;\;\;\;\;kl}=I^{ij}_{\;\;\;kl}.&
  \end{eqnarray}
\end{subequations}
This completes our derivation of the minimum-variance unbiased
estimator $\hat{h}^{ij}$ and its covariance $S^{ij}_{\;\;\;kl}$.  Let
us add a few remarks.

{\bf Remark 1.}  The minimum-variance unbiased estimator for the trace
of $h^{ij}$ is $\hat{h}=\delta_{ij}\hat{h}^{ij}$, with variance
$\langle\hat{h}(f)\hat{h}(f')^{\ast}\rangle-\langle\hat{h}(f)\rangle
\langle\hat{h}(f')^{\ast}\rangle=\delta_{ij}S^{ij}_{\;\;\;kl}(f)\delta^{kl}
\delta(f-f')$.  If the porcupine detects a gravitational wave,
$\hat{h}$ should be consistent with zero, to within this predicted
uncertainty; this is an important observational test to distinguish
genuine gravitational waves from spurious signals.

{\bf Remark 2.}  If $\delta^{ij}\tilde{W}_{ij}^{\alpha}=0$ (as in a
network of equal-arm Michelson interferometers like LIGO/VIRGO
\cite{LIGO}) the porcupine is insensitive to the trace of $h^{ij}$,
and we can only construct an estimator for the traceless part of
$h^{ij}$.  It is convenient to introduce the notation $[\ldots]^{T}$
to denote the traceless part of a tensor.  We must be careful, since
this paper involves the relationship between several different linear
spaces, each with its own trace.  In particular, $[\ldots]^{T}$ will
denote tracelessness with respect to 3-dimensional traces ({\it i.e.}\
those that can be performed using $\delta_{ij}$ and $\delta^{ij}$, but
{\it not} $\delta^{i}_{\;j}$).  Explicitly:
\begin{subequations}
  \begin{eqnarray}
    \big[T^{ij}\big]^{T}\!\!\!&\!=\!&\!\!T^{ij}\!-\!\frac{1}{3}\delta^{ij}\delta_{kl}T^{kl},\\
    \big[T^{ij}_{\;\;\;kl}\big]^{T}\!\!\!&\!=\!&\!\!T^{ij}_{\;\;\;kl}
    \!-\!\frac{1}{3}\delta^{ij}\delta_{mn}T^{mn}_{\;\;\;\;\;kl}
    \!-\!\frac{1}{3}T^{ij}_{\;\;\;mn}\delta^{mn}\delta_{kl}\nonumber\\
    \!&\!\!&\!\!+\frac{1}{9}\delta^{ij}\delta_{pq}T^{pq}_{\;\;\;mn}\delta^{mn}\delta_{kl}.
  \end{eqnarray}
\end{subequations}
Thus, we want the minimum-variance unbiased estimator
$[\hat{h}^{ij}]^{T}$ and its covariance.  This is done by following
the same steps as above; in fact, the answer is still given by
Eqs.~(\ref{def_V} -- \ref{def_D}), as long as we replace each factor
in these equations by its traceless part: {\it e.g.}\
$\hat{h}^{ij}\to[\hat{h}^{ij}]^{T}$,
$\tilde{V}^{ij}_{\;\alpha}\to[\tilde{V}^{ij}_{\;\alpha}]^{T}$,
$I^{ij}_{\;\;\;kl}\to[I^{ij}_{\;\;\;kl}]^{T}$,
$S^{ij}_{\;\;\;kl}\to[S^{ij}_{\;\;\;kl}]^{T}$.  Here
\begin{equation}
  [I^{ij}_{\;\;\;kl}]^{T}=\delta^{[i}_{\;\;[k}\delta^{j]}_{\;\;\,l]}
  -\frac{1}{3}\delta_{}^{ij}\delta^{}_{kl}
\end{equation}
is the identity operator on the 5-dimensional space of $3\times3$
symmetric traceless matrices.  The estimator $[\hat{h}^{ij}]^{T}$ only
exists if $\tilde{W}^{\alpha}_{ij}$ has a left-inverse
$[\tilde{V}^{ij}_{\;\alpha}]^{T}$ in the sense of (the traceless part
of) Eq.~(\ref{VW_constraint}); and this is only possible for $N\geq5$
detectors.

{\bf Remark 3.}  Although the porcupine may be simultaneously immersed
in many gravitational plane waves, travelling in many different
directions, suppose that there is at most one plane wave in each
frequency bin $f\pm\delta f$, traveling in an unknown direction
$\hat{n}(f)$.  Then, since the gravitational wave is transverse, we
have $\tilde{h}^{ij}(f)\hat{n}_{j}(f)=0$, which implies that ${\rm
  Det}[\tilde{h}^{ij}(f)]$ should vanish.  We can check this
observationally by confirming that the estimator ${\rm
  Det}[\hat{h}^{ij}(f)]$ is consistent with zero, to within the
predicted uncertainty.  (Calculating the predicted variance of this
estimator is an exercise in applying Wick's Theorem.)  Also note that
if we choose two mutually orthogonal unit vectors $\hat{a}$ and
$\hat{b}$, and use these to form the estimators
$\vec{A}^{i}=\hat{h}^{ij}\hat{a}_{j}$ and
$\vec{B}^{i}=\hat{h}^{ij}\hat{b}_{j}$, then
$\hat{N}=(\vec{A}\times\vec{B})/\big|\vec{A}\times\vec{B}\big|$ is an
estimator for the zero-eigenvector of $h_{ij}$ ({\it i.e.}\
$\pm\hat{n}$, the propagation direction up to a sign).

\section{Average SNR$^{2}$}

Let us average the expected SNR$^{2}$ (\ref{SNR}): first over the
polarization state, and then also over the direction $\hat{n}$ of the
incident gravitational wave.  First we average over the polarization,
keeping $\hat{n}$ fixed, to obtain:
\begin{equation}
  {\rm SNR}_{\hat{n}}^{2}=
  \int_{-\infty}^{\infty}df\,\big|\tilde{h}(f)\big|^{2}
  {\cal P}^{kl}_{\;\;\;ij}(\hat{n})K^{ij}_{\;\;\;kl}(f)
\end{equation}
where 
\begin{subequations}
  \begin{eqnarray}
    \big|\tilde{h}(f)\big|^{2}\!&\!\equiv\!&\!\tilde{h}_{ij}^{\ast}(f)\tilde{h}^{ij}(f), \\
    {\cal P}^{ij}_{\;\;\;kl}(\hat{n})
    \!&\!\equiv\!&\!(1/2)\delta^{A}_{\;\;B}
    P^{ij}_{A}(\hat{n})P_{kl}^{B}(\hat{n})^{\ast} \\
    \!&\!=\!&\!(1/2)\Delta^{[i}_{\;\;[k}\Delta^{j]}_{\;\;\,l]}-(1/4)\Delta^{ij}\Delta_{kl},
  \end{eqnarray}
\end{subequations}
and $\Delta_{ij}\equiv\delta_{ij}-\hat{n}_{i}\hat{n}_{j}$ is the
projector onto the plane perpendicular to $\hat{n}$.  Now use
Eq.~(2.3b) in \cite{Thorne} to find
\begin{equation}
  \int \frac{d\Omega}{4\pi}{\cal P}^{ij}_{\;\;\;kl}(\hat{n})
  =\frac{1}{5}[I^{ij}_{\;\;\;kl}]^{T}
\end{equation}
and use this identity to obtain the SNR$^{2}$, fully averaged over
both the polarization and direction $\hat{n}$ of the the incident
gravitational wave:
\begin{equation}
  \label{SNR_average}
   \langle{\rm SNR}^{2}\rangle=
   \int_{-\infty}^{\infty}df\big|\tilde{h}(f)\big|^{2}
   \frac{1}{5}[I^{kl}_{\;\;\;ij}]^{T}K^{ij}_{\;\;\;kl}(f). 
\end{equation}
%If the noise is uncorrelated between different detectors,
%$S^{\alpha}_{\;\;\beta}(f)=\delta^{\alpha}_{\;\;\beta}S_{\beta}(f)$,
%then
%\begin{equation}
%  \langle{\rm SNR}^{2}\rangle\!=\!\!\int_{-\infty}^{\infty}
%  \!\!\!df\big|\tilde{h}(f)\big|^{2}\sum_{\alpha}
%  \frac{[I^{kl}_{\;\;\;ij}]^{T}\tilde{W}^{ij}_{\alpha}(f)^{\ast}
%    \tilde{W}^{\alpha}_{kl}(f)}{5S_{\alpha}(f)}.
%\end{equation}

\section{Perfect Porcupines}

A perfect porcupine is a network for which the tensor $K^{ij}_{\;\;\;kl}(f)$ is
rotationally invariant.  The only objects from which we can build such
a tensor are $\delta^{ij}$ and $\epsilon^{ijk}$.  Imposing the
constraints (\ref{K_constraints}) we find that the most general
perfect porcupine is of the form
\begin{equation}
  \label{K_perfect_porcupine}
  K^{ij}_{\;\;\;kl}(f)=F(f)[I^{ij}_{\;\;\;kl}]^{T}+G(f)\delta^{ij}\delta_{kl}
\end{equation}
where $F$ and $G$ are real and symmetric under $f\to-f$
\begin{equation}
  F(f)\!=\!F(f)^{\ast}\!=\!F(-f),\quad G(f)\!=\!G(f)^{\ast}\!=\!G(-f).
\end{equation}
Then the expressions for the SNR$^{2}$, and the angular parts of the
Fisher matrix, simplify beautifully
\begin{subequations}
  \begin{eqnarray}
    \label{SNR_perfect_porcupine}
    &{\rm SNR}^{2}=\int_{-\infty}^{\infty}df\,F(f)
    \big|\tilde{h}(f)\big|^{2},& \\
    &\Gamma_{\bar{\mu}\bar{\nu}}=({\rm
      SNR}^{2})\delta_{\bar{\mu}\bar{\nu}},\qquad
    \Gamma_{\bar{\mu}\chi}=\Gamma_{\chi\bar{\mu}}=0.&
  \end{eqnarray}
\end{subequations}
Thus, a perfect porcupine localizes a gravitational wave source within 
a circular spot (really two antipodal spots) of radius $1/{\rm SNR}$, 
regardless of the source's properties.

\section{Simple Porcupines}

Suppose that the individual detectors in the network are identical to
one another (up to spatial orientation), with noise that is
uncorrelated between different detectors: $S^{\alpha}_{\;\;\beta}(f)
=S(f)\delta^{\alpha}_{\;\;\beta}$.  Further suppose that each
detector's frequency response may be factored out from its tensor
structure: $\tilde{W}_{ij}^{\alpha}(f)=\tilde{W}(f) A_{ij}^{\alpha}$.
(We can take the frequency-independent matrices $A_{ij}^{\alpha}$ to
be normalized: $A^{ij\ast}_{\alpha}A_{ij}^{\alpha}=1$ with $\alpha$
unsummed.)  A network with these properties will be called ``simple.''

The averaged SNR$^{2}$ (\ref{SNR_average}) for a simple porcupine is
\begin{equation}
  \label{SNR_average_factored}
  \langle{\rm SNR}^{2}\rangle=N\frac{(3-\big|{\rm Tr}\,A\big|^{2})}{15}
  \int_{-\infty}^{\infty}df\frac{\big|\tilde{h}(f)\big|^{2}\big|\tilde{W}(f)\big|^{2}}{S(f)}
\end{equation}
where ${\rm Tr}\,A=\delta^{ij}A_{ij}^{\alpha}$.

A ``simply perfect porcupine'' is both simple and perfect.  In this
case, we can trace Eq.~(\ref{K_perfect_porcupine}) in two inequivalent
ways (by contracting with $\delta^{k}_{\;\;i} \delta^{l}_{\;\;j}$ or
$\delta_{ij}\delta^{kl}$) to find
\begin{subequations}
  \begin{eqnarray}
    F(f)&=&N\frac{\big|\tilde{W}(f)|^{2}}{S(f)}
    \frac{(3-\big|{\rm Tr}\,A\big|^{2})}{15}, \\
    G(f)&=&N\frac{\big|\tilde{W}(f)|^{2}}{S(f)}
    \frac{\big|{\rm Tr}\,A\big|^{2}}{9}.
 \end{eqnarray}
\end{subequations}
For such a porcupine, we see that the SNR$^{2}$ expression
(\ref{SNR_perfect_porcupine}) coincides with the averaged expression
(\ref{SNR_average_factored}) for a simple but imperfect porcupine.
Note that these expression depend {\it only} on $N$, $S(f)$,
$\tilde{W}(f)$, and ${\rm Tr}\,A$; and {\it not} on the directional
orientation of the individual detectors that make up the network.  In
other words, if two different simply perfect porcupines are built from
the same collection ({\it i.e.}\ the same number and type) of
individual detectors, then they will have identical properties,
despite the fact that they may seem like entirely different network
configurations.

Now consider the problem of finding simply perfect porcupine
configurations explicitly.  Two cases deserve particular attention.
In the first case, the network is constructed from single-arm
detectors (like the proposed AGIS detector \cite{DimopoulosEtAl,
  YuTinto, HohenseeEtAl}); the arm of detector $\alpha$ lies along the
direction $\hat{k}_{\alpha}$ so that
$A_{ij}^{\alpha}=\hat{k}_{i}^{\alpha}\hat{k}_{j}^{\alpha}$ and ${\rm
  Tr}\,A=1$.  In the second case, the network is constructed from
double-arm detectors (like the LIGO/VIRGO Michelson interferometers
\cite{LIGO}); the two arms of detector $\alpha$ lie along the two
perpendicular directions $\hat{p}_{\alpha}$ and $\hat{q}_{\alpha}$ so
that $A_{ij}^{\alpha}=(\hat{p}_{i}^{\alpha}\hat{p}_{j}^{\alpha}-
\hat{q}_{i}^{\alpha}\hat{q}_{j}^{\alpha})/\sqrt{2}$ and ${\rm
  Tr}\,A=0$.  In these two cases, Eq.~(\ref{K_perfect_porcupine})
becomes
\begin{subequations}
  \begin{eqnarray}
    \label{single_arm}
    &\sum_{\alpha}\hat{k}_{\alpha}^{i}\hat{k}_{\alpha}^{j}
    \hat{k}_{\alpha}^{k}\hat{k}_{\alpha}^{l}\!=\!
    (N/15)(\delta^{ij}\delta^{kl}\!+\!\delta^{ik}\delta^{jl}\!+\!\delta^{il}\delta^{jk}),& \\
    \label{double_arm}
    &\sum_{\alpha}(\hat{p}_{\alpha}^{i}\hat{p}_{\alpha}^{j}
    \!-\!\hat{q}_{\alpha}^{i}\hat{q}_{\alpha}^{j})
    (\hat{p}^{\alpha}_{k}\hat{p}^{\alpha}_{l}\!-\!
    \hat{q}^{\alpha}_{k}\hat{q}^{\alpha}_{l})\!=\!(2N/5)[I^{ij}_{\;\;\;kl}]^{T}.&\qquad
  \end{eqnarray}
\end{subequations}
Note that both sides of the single-arm Eq.~(\ref{single_arm}) are
completely symmetric under permutation of the indices $\{i,j,k,l\}$;
this gets us down to 15 independent equations, one of which (obtained
by tracing both sides with $\delta_{ij}\delta_{kl}$) is automatically
satisfied.  Thus we have 14 independent equation for $2N-3$ variables
(2 angles for each $\hat{k}_{\alpha}$, minus 3 angles corresponding to
an arbitrary rigid rotation of the entire network).  So, for $N\geq9$,
Eq.~(\ref{single_arm}) will have a $(2N-17)$ parameter family of
inequivalent solutions.  In addition, there is a unique smaller
solution with $N=6$ detectors along the 6 diameters of an icosahedron
\cite{JohnsonMerkowitz, CerdonioEtAl, porcupine}.  On the other hand,
for the double-arm Eq.~(\ref{double_arm}), there are no solutions for
$N\leq4$, and continuous families of inequivalent solutions for all
$N\geq5$.

\section{Discussion}

Let us highlight three open problems.  (i) The first problem is to
find the general solution of Eq.~(\ref{single_arm}) or
Eq.~(\ref{double_arm}) [that is, the most general configuration of a
(single-arm or double-arm) simply perfect porcupine].  Which among
these configurations could be most practically situation on the
Earth's available land area?  (ii) Whereas perfect porcupines have a
completely isotropic sensitivity pattern, ``pointed porcupines'' lie
at the opposite extreme: they are the porcupines with the sensitivity
pattern that is most highly localized on the sky.  The second problem
would be to develop the theory of such pointed porcupines,
particularly since they may be the configurations that maximize the
expected event rate (for a fixed collection of detectors, and assuming
the gravitational wave sources are Poisson distributed throughout
space).  The third problem is to extend the present formalism in a
relativistically correct way to investigate networks in which each
detector is allowed to follow its own (possibly accelerating and
rotating) trajectory.

{\bf Acknowledgements.}  I am very grateful to Mike Kesden and Paul
McFadden for valuable conversations.  This work was supported in part
by the CIFAR JFA.

\end{document}